\documentclass[a4paper,11pt,dvips]{article}
\usepackage{graphicx}
\usepackage{amsmath}
\usepackage{amssymb}
\usepackage{latexsym}
\usepackage{subcaption}
\usepackage[colorlinks]{hyperref}
\usepackage{color}
\usepackage{enumerate}
\usepackage{enumitem}
\usepackage{cite}
\usepackage{makeidx}
\usepackage{colortbl}

\newcommand{\bra}{\begin{array}}
\newcommand{\era}{\end{array}}
\newcommand{\beq}{\begin{equation}}
\newcommand{\eeq}{\end{equation}}
\newcommand{\beqar}{\begin{eqnarray}}
\newcommand{\eeqar}{\end{eqnarray}}

\def\BC{\bb C}
\def\_\BC{\bbi C}

\def\( {\left(}
   \def\) {\right)}
\def\[ {\left[}
\def\] {\right]}
\def\no2 {{\textstyle{n\over 2}}}

\def\dag {{\dagger}}

\newcommand{\lb}{\label}

\begin{document}
\begin{titlepage}
\setcounter{page}{1}
\renewcommand{\thefootnote}{\fnsymbol{footnote}}

\begin{flushright}
\end{flushright}

\vspace{5mm}
\begin{center}

{\Large \bf {
AA-stacked Bilayer Graphene
Quantum Dots in 
Magnetic Field}}

\vspace{5mm}

{\bf Abdelhadi Belouad}$^{a}$, {\bf Youness Zahidi}$^{a}$ and {\bf Ahmed Jellal\footnote{\sf
ajellal@ictp.it -- jellal.a@ucd.ac.ma}}$^{a,b}$

\vspace{5mm}

{$^{a}$\em Theoretical Physics Group,  
Faculty of Sciences, Choua\"ib Doukkali University},\\
{\em PO Box 20, 24000 El Jadida,
Morocco}

{$^b$\em Saudi Center for Theoretical Physics, Dhahran, Saudi Arabia}

\vspace{10mm}

{\sf Dedicated to Prof. Dr. Hachim A. Yamani on the occasion of his 70th birthday}


\vspace{3cm}

\begin{abstract}

By applying the infinite-mass boundary condition, we analytically
calculate the confined states and the corresponding wave functions
of AA-stacked bilayer graphene quantum {dots}
 in the presence of an uniform magnetic field $B$.
 It is found that
the energy spectrum shows two set of levels, which
are the double copies of the energy spectrum for
single layer graphene, shifted up-down  by $+\gamma$ and
$-\gamma$, respectively. However, the obtained spectrum exhibits
different symmetries between the electron and hole states
as well as the intervalley symmetries. It is noticed that, the applied
magnetic field breaks all symmetries, except
one related to the intervalley electron-hole symmetry,
i.e. $E^e(\tau,m)=-E^h(\tau,m)$.
Two different regimes of 
confinement are found:
the first one is due to the infinite-mass barrier at weak $B$ and the
second is dominated by the magnetic field as long as $B$ is large.
We numerically investigated the basics features of the energy spectrum to show the main
similarities and differences with respect to
monolayer graphene, AB-stacked bilayer graphene and
semiconductor quantum dots.




\end{abstract}
\end{center}

\vspace{5cm}

\noindent PACS numbers: 81.05.ue, 81.07.Ta, 73.22.Pr\\
 \noindent Keywords:
AA-stacked bilayer graphene, quantum dot, infinite-mass, magnetic
field.

\end{titlepage}


\section{Introduction}

Graphene~\cite{NGMJZDGF04,NJBKMG05} is the name given to the
atomically thin layer of carbon atoms that can be viewed either as
a single layer of graphite or an unrolled nanotube. The
experimental breakthrough~\cite{NGMJZDGF04}, has generated much
excitement in the condensed matter physics community. The large
interest is due to both the unusual mechanical and electronic
properties as well as for the prospects of applications, which may lead
to their use in novel nanoelectronic devices. Graphene can provide
a good platform for the study of the electronic properties of a
pure two-dimensional system. In fact, graphene has an unique band
structure, which is gapless and exhibits a linear dispersion
relation at two inequivalent points, labeled as $K$ and $K'$ in
the Brillouin zone. In addition, graphene
presents a variety of exotic electronic properties like
electron-hole symmetry~\cite{CGPN09}, Klein tunneling~\cite{KNG06}
and anomalous quantum Hall effect~\cite{GS05}. The equation
describing the electronic excitations in graphene is formally
similar to the Dirac equation for massless fermions, which travel
at a speed of the order of $v_F\approx10^6$ $m/s$~\cite{S84,DM84}.

Graphene can not only exist in the free state, but two or more
layers can stack above each other to form what is called few layer
graphene. As example the bilayer graphene (BLG) that is the stack
of two sheets.
There are two dominant ways in which the two layers can be
stacked. The first one is the so called AB-stacked
BLG~\cite{B24,CGM91} and the second is the AA-stacked
BLG~\cite{LLAK08,ARV08}. In the case of AB-stacked BLG, only one
atom in the lower layer lies directly below an other atom in the
upper layer and the other two atoms over the center of the hexagon
in the other layer. However, for AA-stacked BLG the A sublattice
of the top layer is stacked directly above the same sublattice of
the bottom layer. The BLG exhibits additional properties that can
not be found
in single layer graphene.
Indeed, the AB-stacked BLG has a gapless quadratic dispersion
relation, two conduction bands and two valance bands, each pair is
separated by an interlayer coupling energy of order $\gamma_1 =
400\ meV$. However, the energy bands for AA-stacked BLG are just
the double copies of single layer graphene bands shifted up and
down by the interlayer coupling $\gamma = 200\ meV$.

Graphene quantum dots (QDs)~\cite{C99,SE07,MP08} have sparked
intense research activities related to quantum information storage
and realization of a system based on the QD.
The Klein tunneling effect~\cite{KNG06} in graphene prevents
carrier confinement and thus prevents the realization of QDs in
graphene~\cite{SE07,MP08}. From an application technological point
of view, the confinement of electrons is of particular importance
to build functional nano-devices. Hence, manufacturing
graphene based electronic devices will be a great
challenge~\cite{PMPV06,MP08}. Several studies showed that the
confinement of carriers can be
made whether, 
through an external magnetic field~\cite{GMR09} or finite mass
term with an electrostatic potential~\cite{RNBT09}.
Recently, the energy levels of circular graphene QDs in the
presence and absence of a external magnetic field, was
investigated, for the infinite-mass and zigzag boundary
conditions~\cite{GZCTFP11}. Moreover, theoretical studies were
reported
on circular, triangular and hexagonal graphene QDs for different
boundary conditions in the presence of a perpendicular magnetic
field~\cite{DMAGF14,ZCFP11,RN10,SES08}.

In this paper, we
consider a circular QD in AA-stacked bilayer graphene in the
presence of an external magnetic field.
We solve the Dirac equation in the barrier region (region II) and
apply the boundary conditions, for the four wave
function components on the solid circles shown in
Figure~\ref{barrier} for the limit $V\longrightarrow \infty$, to end up with
the infinite mass boundary condition. For this, we
start with the system shown in Figure~\ref{barrier}(a) and match 
the wave function components in regions I and III as well as make use of the
appropriate approximations. 
Subsequently,
we obtain the energy levels for circular QDs in AA-stacked BLG corresponding to both valleys $K$
and $K'$ in terms of the dot
radius and the magnetic field. We investigate the basic features of  our results and compare them with
those for circular QDs in monolayer graphene and AB-stacked
BLG. 

The present paper is organized as follows. In section $2$, we formulate our problem by setting the theoretical
model that is relevant to describe our system. Subsequently, we use  
the eigenvalue equation 
to explicitly determine
the infinite mass boundary conditions. This will be employed, in section 3, to separately study  the energy spectrum of
our system in the 
absence and presence of the magnetic field.
To underline the behavior of our system
we numerically investigate the basic features of
the obtained results 
 in section $4$. In the final section, we conclude our results.

\section{Problem setting}

\begin{figure}[!ht]
\centering
\includegraphics[width=6cm, height=5cm]{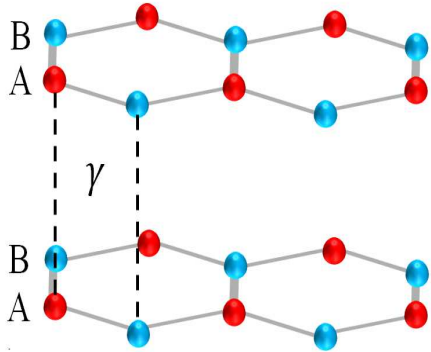}
\caption{ {\sf {Schematic illustration of lattice structure of
AA-stacked bilayer graphene. It is consists of two graphene
layers. Each carbon atom of the upper layer is located above the
corresponding atom of the lower layer and they are separated by an
interlayer coupling energy $\gamma$. The unit cell of the
AA-stacked bilayer graphene consists of four atoms A, B, A' and
B'.}}}\label{struc}
\end{figure}

To set our problem, let us  
consider a bilayer 
graphene system consists of two monolayers,
arranged
in the Bernal AA stacking
way. Each layer consists of two triangular sublattices,
labeled A and B (upper layer) and A$^{'}$ and B$^{'}$ (lower
layer). The A sublattice of the upper layer is stacked directly
above the same sublattice of the lower layer by the interlayer
coupling $\gamma=200\ meV$~\cite{CE12}  as presented in Figure~\ref{struc}.

\begin{figure}[!ht]
        \centering
        \begin{subfigure}[b]{0.3\textwidth}
                \includegraphics[width=5cm, height=5cm]{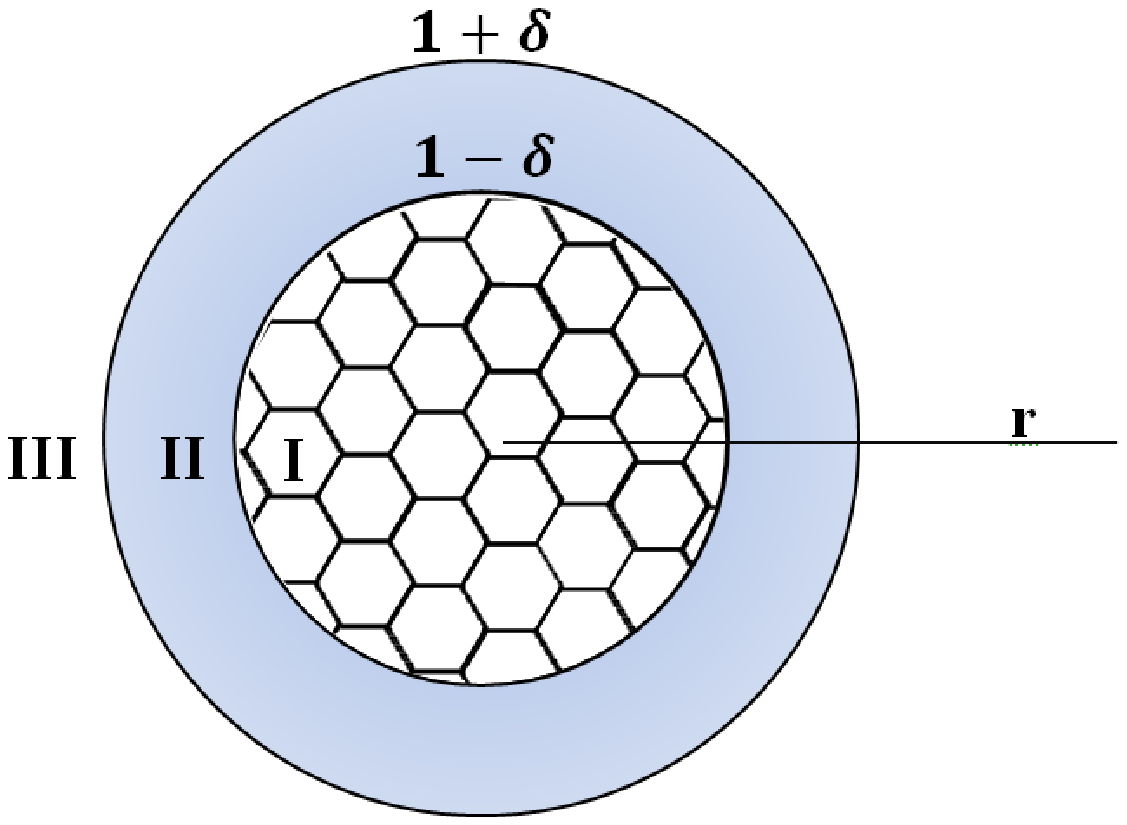}
                \caption{}
                \label{fig:c1}
        \end{subfigure}%
        \begin{subfigure}[b]{0.3\textwidth}
                \includegraphics[width=5cm, height=5cm]{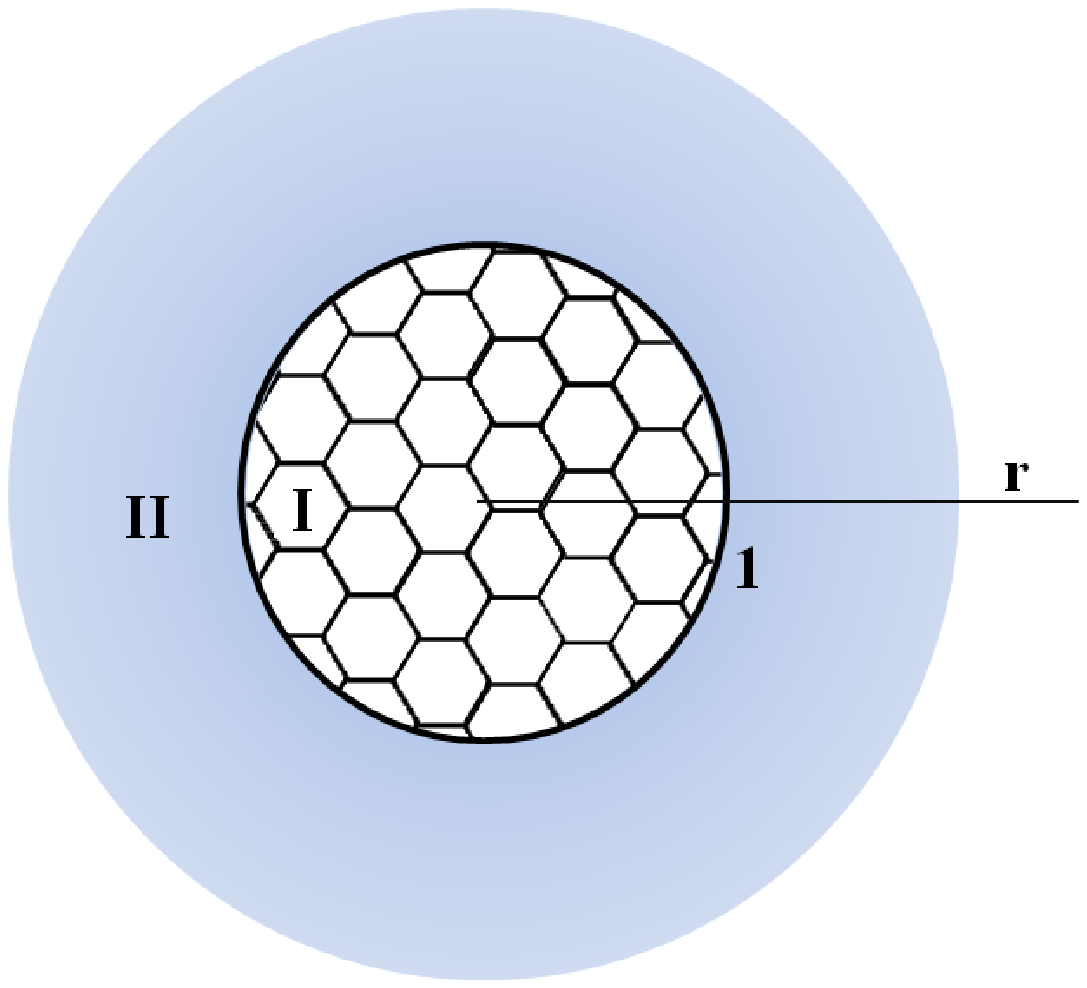}
                \caption{}
                \label{fig:a1}
        \end{subfigure}
        \caption{\sf{ Illustration of the mass barrier profile. (\subref{fig:c1}): a ring-shaped barrier. (\subref{fig:a1}): circular
quantum dot, made of AA-stacked bilayer graphene surrounded by an
infinite-mass barrier.
}}\label{barrier}
\end{figure}

Next, 
we assume that the carriers in AA-stacked bilayer graphene are confined in a circular area of
radius $R$, which can be  modeled by an infinite-mass barrier. For this, we will
derive the infinite-mass boundary conditions for the Dirac electron
interacting with circular barrier structures as shown in blue
color 
(Figure~\ref{barrier}),
in the presence of a perpendicular magnetic field. More precisely, Figure~\ref{barrier} 
illustrates a schematic depiction of the mass
barrier profile for an AA-stacked BLG quantum dots that will be used 
to deal  with different issues.

To be much more concrete,
we consider only nearest neighbor in-plane hopping, as well as
direct interlayer hopping, i.e. A to A$^{'}$ or B to B$^{'}$
sublattices. Then, in the basis of $\Psi=\left(\Psi_{A},\ \Psi_{B},\
\Psi_{B^{'}},\ \Psi_{A^{'}}\right)^T$, where $\Psi_{A\ (A^{'})}$
and $\Psi_{B\ (B^{'})}$ are the envelope functions associated with
the probability amplitudes of the wave functions on the A
(A$^{'}$) and B (B$^{'}$) sublattices of the upper (lower)
layer, the Hamiltonian describing our system is given by
\beq \label{eq42}
H =
\begin{pmatrix}
\tau V(r) & \pi & 0 & \gamma \\
\pi^\dag & -\tau V(r) & \gamma & 0 \\
0 & \gamma &  \tau V(r) & \pi^\dag  \\
\gamma & 0 & \pi & -\tau V(r)
\end{pmatrix}
\eeq
where the momentum operators in polar
coordinates
\beq
\pi=-i\hbar v_Fe^{i\theta}\left(\frac{\partial}
{\partial r}+\frac{i}{r}\frac{\partial}{\partial \theta}\right),\qquad
\pi^\dag =-i\hbar v_Fe^{-i\theta}\left(\frac{\partial}
{\partial r}-\frac{i}{r}\frac{\partial}{\partial \theta}\right)
\eeq
$v_F\approx 10^6 m/s$ is the Fermi velocity and
$\tau=\pm$ differentiates the two valleys $K$ and $K'$. The mass
profile $V(r)$ is considered to be circularly symmetric
\beq \label{Vform} V(r) =  \left \{ \begin{array}{cc} 0 ,& \;r<R
\;
\\ \infty, &  \; r>R \;
\end{array} \right.
\eeq
and $r$ is the radial coordinate of the polar coordinates
and $R$ is the radius of the quantum dot. Since
the total angular momentum operator $J_z$ commutes with $H$, then we can
construct a common basis in terms of the eigenspinors, such as 
\beq\label{eigen}
\Psi(r,\theta)=\begin{pmatrix}
\Phi_A(r)e^{im\theta} \\
i\Phi_B(r)e^{i(m-1)\theta}\\
i\Phi_{B'}(r)e^{i(m-1)\theta}\\
\Phi_{A'}e^{im\theta}
\end{pmatrix}
\eeq 
where $m$ is the angular momentum quantum number, which being an
integer.
It is convenient to switch to the dimensionless variables by
setting the quantities
\beq
\frac{r}{R}\longrightarrow r, \qquad
\frac{R}{\hbar v_F}E \longrightarrow E, \qquad \frac{R}{\hbar
v_F}\gamma \longrightarrow \gamma, \qquad \frac{R}{\hbar v_F} V
\longrightarrow V.
\eeq
Plugging
\eqref{eigen} into the eigenvalue equation
$H\Psi(r,\theta)=E\Psi(r,\theta)$, to obtain the following coupled
differential equations
\begin{subequations}\lb{DE3}
\beqar
&&\left(\frac{\partial}{\partial r}+\frac{m}{r}\right)\Phi_{A}(r) = -(E+\tau V)\Phi_{B}(r)+ \gamma\Phi_{B'} (r)\\
&&\left(\frac{\partial}{\partial r}-\frac{m-1}{r}\right)\Phi_{B}(r) = (E-\tau V)\Phi_{A}(r)- \gamma\Phi_{A'}(r)\\
&&\left(\frac{\partial}{\partial
r}-\frac{m-1}{r}\right)\Phi_{B'}(r)=(E+\tau V)\Phi_{A'}(r)-
\gamma\Phi_{A}(r)\\
&&\left(\frac{\partial}{\partial r}+\frac{m}{r}\right)\Phi_{A'}(r)
=-(E-\tau V)\Phi_{B'}(r)+ \gamma\Phi_{B}(r).
 \eeqar
\end{subequations}
Before getting the solutions of the energy spectrum,
we first look for the infinite-mass boundary condition for AA-stacked
bilayer graphene. This can be done
by solving the differential equations~\eqref{DE3} in the barrier
region (region II). Indeed,
we consider the ring-shaped barrier shown in
Figure~\ref{barrier}(\subref{fig:c1}) and
take the limit $V \longrightarrow \infty$ to end up with the
approximate set of differential equations
\begin{subequations}\lb{eq48}
\beqar
&&\frac{\partial}{\partial r}\Phi_{A}(r)=  -\tau V\Phi_{B}(r)+ \gamma\Phi_{B'}(r)\\
&&\frac{\partial}{\partial r}\Phi_{B}(r)= -\tau V \Phi_{A}(r)- \gamma\Phi_{A'}(r) \\
&&\frac{\partial}{\partial r}\Phi_{B'}(r)= \tau V \Phi_{A'} (r)-
\gamma \Phi_{A}(r)\\
&&\frac{\partial}{\partial r}\Phi_{A'}(r)= \tau V\Phi_{B'}(r)+
\gamma\Phi_{B}(r).
 \eeqar
\end{subequations}
These give the eigenspinor
solution in the thin shadowed region II, delimited by
two circles of radius $1\pm \delta$ with the condition $\delta \ll
1$, such as
\begin{subequations}\lb{eq50}
\beqar
&&\Phi_{A}(r)=a e^{(r-1)\alpha_+}+ b e^{-(r-1)\alpha_+}+c e^{(r-1)\alpha_-}+ d e^{-(r-1)\alpha_-} \\
&&\Phi_{B}(r)=\frac{-\tau}{V}\left(\alpha_+a
e^{(r-1)\alpha_+}-\alpha_+b e^{-(r-1)\alpha_+}+\alpha_-c
e^{(r-1)\alpha_-}-\alpha_-d e^{-(r-1)\alpha_-}\right) \\
&&\Phi_{B'}(r)=\frac{i}{V}\left(\alpha_+a e^{(r-1)\alpha_+}-\alpha_+be^{-(r-1)\alpha_+}-\alpha_-c e^{(r-1)\alpha_-}+\alpha_-d e^{-(r-1)\alpha_-}\right) \\
&&\Phi_{A'}(r)=i\tau\left(a e^{(r-1)\alpha_+}+
be^{-(r-1)\alpha_+}-c e^{(r-1)\alpha_-}-d
e^{-(r-1)\alpha_-}\right) \eeqar
\end{subequations}
where the parameter $\alpha$ is defined by
\beq
\alpha^2_\pm=(V\pm i\gamma)^2
\eeq
and the coefficients ($a,b,c,d$) are constants.
Using the continuity of the wave functions at the boundary
conditions on the two circles
to obtain the solutions at the boundary
 $1-\delta$
\begin{subequations}\lb{eq51}
\beqar
&&\Phi^{\text{I}}_{A}(1-\delta)=a e^{-\delta\alpha_+}+b e^{\delta\alpha_+}+c e^{-\delta\alpha_-}+d e^{\delta\alpha_-} \\
&&\Phi^{\text{I}}_{B}(1-\delta)=\frac{-\tau}{V}\left(\alpha_+a e^{-\delta\alpha_+}-\alpha_+b e^{\delta\alpha_+}+\alpha_-c e^{-\delta\alpha_-}-\alpha_-d e^{\delta\alpha_-}\right) \\
&&\Phi^{\text{I}}_{B'}(1-\delta)=\frac{i}{V}\left(\alpha_+a e^{-\delta\alpha_+}-\alpha_+b e^{\delta\alpha_+}-\alpha_-c e^{-\delta\alpha_-}+\alpha_- de^{\delta\alpha_-}\right) \\
&&\Phi^{\text{I}}_{A'}(1-\delta)=i\tau\left(a
e^{-\delta\alpha_+}+b e^{\delta\alpha_+}-c e^{-\delta\alpha_-}-d
e^{\delta\alpha_-}\right) \eeqar
\end{subequations}
as well as at $1+\delta$
\begin{subequations}\lb{eq52}
\beqar
&&\Phi^{\text{III}}_{A}(1+\delta)=a e^{\delta\alpha_+}+b e^{-\delta\alpha_+}+c e^{\delta\alpha_-}+d e^{-\delta\alpha_-} \\
&&\Phi^{\text{III}}_{B}(1+\delta)=\frac{-\tau}{V}\left(\alpha_+a e^{\delta\alpha_+}-\alpha_+b e^{-\delta\alpha_+}+\alpha_-c e^{\delta\alpha_-}-\alpha_-d e^{-\delta\alpha_-}\right)\\
&&\Phi^{\text{III}}_{B'}(1+\delta)=\frac{i}{V}\left(\alpha_+a e^{\delta\alpha_+}-\alpha_+b e^{-\delta\alpha_+}-\alpha_-c e^{\delta\alpha_-}+\alpha_-d e^{-\delta\alpha_-}\right)\\
&&\Phi^{\text{III}}_{A'}(1+\delta)=i\tau\left(a
e^{\delta\alpha_+}+b e^{-\delta\alpha_+}-c e^{\delta\alpha_-}-d
e^{-\delta\alpha_-}\right). \eeqar
\end{subequations}

At the boundary on the two circles demarcating the region II,
the four-component wave-functions should be continuous. Connecting
the wave-function components in regions I and III and taking into
account
$\delta\ll 1$
to write
\beqar\label{eq53}
\Phi^{\text{III}}_{B'}(1)-\Phi^{\text{I}}_{B'}(1) &=&-i\tau\left(\Phi^{\text{III}}_{B}(1)-\Phi^{\text{I}}_{B}(1)\right)\\
&& + \frac{\tau\alpha_-}{V}\tanh(\delta
\alpha_-)\left[
\Phi^{\text{III}}_{A'}(1)+\Phi^{\text{I}}_{A'}(1)-i\tau\left(\Phi^{\text{III}}_{A}(1)+\Phi^{\text{I}}_{A}(1)\right)\right] \nonumber\\
\label{eq54}
\Phi^{\text{III}}_{A'}(1)-\Phi^{\text{I}}_{A'}(1) &=&-i\tau\left(\Phi^{\text{III}}_{A}(1)-\Phi^{\text{I}}_{A}(1)\right)\\
&&-\frac{V}{\alpha_+}\tanh(\delta\alpha_+)
\left[i\left(\Phi^{\text{III}}_{B}(1)+\Phi^{\text{I}}_{B}(1)\right)-\tau(\Phi^{\text{III}}_{B'}(1)+\Phi^{\text{I}}_{B'}(1))\right] \nonumber.
\eeqar
To find the general form of the boundary conditions corresponding
to the ring-shaped barrier shown in
Figure~\ref{barrier}(\subref{fig:c1}), we consider the limits
$\delta\longrightarrow 0,V\gg \gamma$ and $\tanh(\delta
\alpha_{\pm})=q$, where
$q$ is a function of the height of the barrier $V$. Thus \eqref{eq53} and \eqref{eq54} become 
\beqar \label{eq55}
&&\Phi^{\text{III}}_{B'}(1)-\Phi^{\text{I}}_{B'}(1)=-i\tau\left(\Phi^{\text{III}}_{B}(1)-\Phi^{\text{I}}_{B}(1)\right)+q
\left[
\tau(\Phi^{\text{III}}_{A'}(1)+\Phi^{\text{I}}_{A'}(1))-i\left(\Phi^{\text{III}}_{A}(1)+\Phi^{\text{I}}_{A}(1)\right)\right]\\
\label{eq56}
&&\Phi^{\text{III}}_{A'}(1)-\Phi^{\text{I}}_{A'}(1)=-i\tau\left(\Phi^{\text{III}}_{A}(1)-\Phi^{\text{I}}_{A}(1)\right)+q
\left[\tau
(\Phi^{\text{III}}_{B'}(1)+\Phi^{\text{I}}_{B'}(1))-i\left(\Phi^{\text{III}}_{B}(1)+\Phi^{\text{I}}_{B}(1)\right)\right].
\eeqar
We can also derive the boundary condition for the structure shown
in Figure~\ref{barrier}(\subref{fig:a1}). Indeed,
we consider the
limits: $V\longrightarrow \infty$, $q=1$ and
$\Phi^{\text{III}}_{A'}(1)=\Phi^{\text{III}}_{B'}(1)=\Phi^{\text{III}}_{A}(1)=\Phi^{\text{III}}_{B}(1)=0$,
to end up with the boundary conditions for an AA-stacked bilayer
quantum dot surrounded by an infinite-mass barrier
\begin{subequations}\lb{IMBC}
\beqar
&&\Phi^{\text{I}}_{B}(1)- \tau \Phi^{\text{I}}_{A}(1)=0 \lb{eq4a}  \\
&& \Phi^{\text{I}}_{B'}(1)+ \tau \Phi^{\text{I}}_{A'}(1)=0
\lb{eq4b}. \eeqar
\end{subequations}
It is important to note that the infinite-mass boundary condition
\eqref{IMBC} is depending only to the  region~I, which means that the region
outside the dot (region II) is forbidden for
particles~\cite{DM84}. In addition,
(\ref{IMBC}a) and~(\ref{IMBC}b) have the same structure except for
a sign, each one of them 
connects the value of the
pseudospin components at the boundary of the two sublattices of
each layer separately. In contrast of those obtained for
AB-stacked bilayer graphene, were they connect each layer to the
other\cite{DMAGF14}. We observe also that each one of these two
conditions ((\ref{IMBC}a) and (\ref{IMBC}b)) has the same form to
that obtained in the case of monolayer graphene~\cite{GZCTFP11}.
Note that from both conditions
we find the energy spectrum, which will be discussed in the next
section.

\section{Energy spectrum}

We will look for the solutions of the energy spectrum inside the quantum
dot by using  the obtained infinite-mass boundary conditions. In doing so, we separately consider the cases of 
the absence and presence of a magnetic
field.

\subsection{Zero magnetic field}

\begin{figure}[h!]
  \centering
 \includegraphics[width=4.3cm, height=4.2cm]{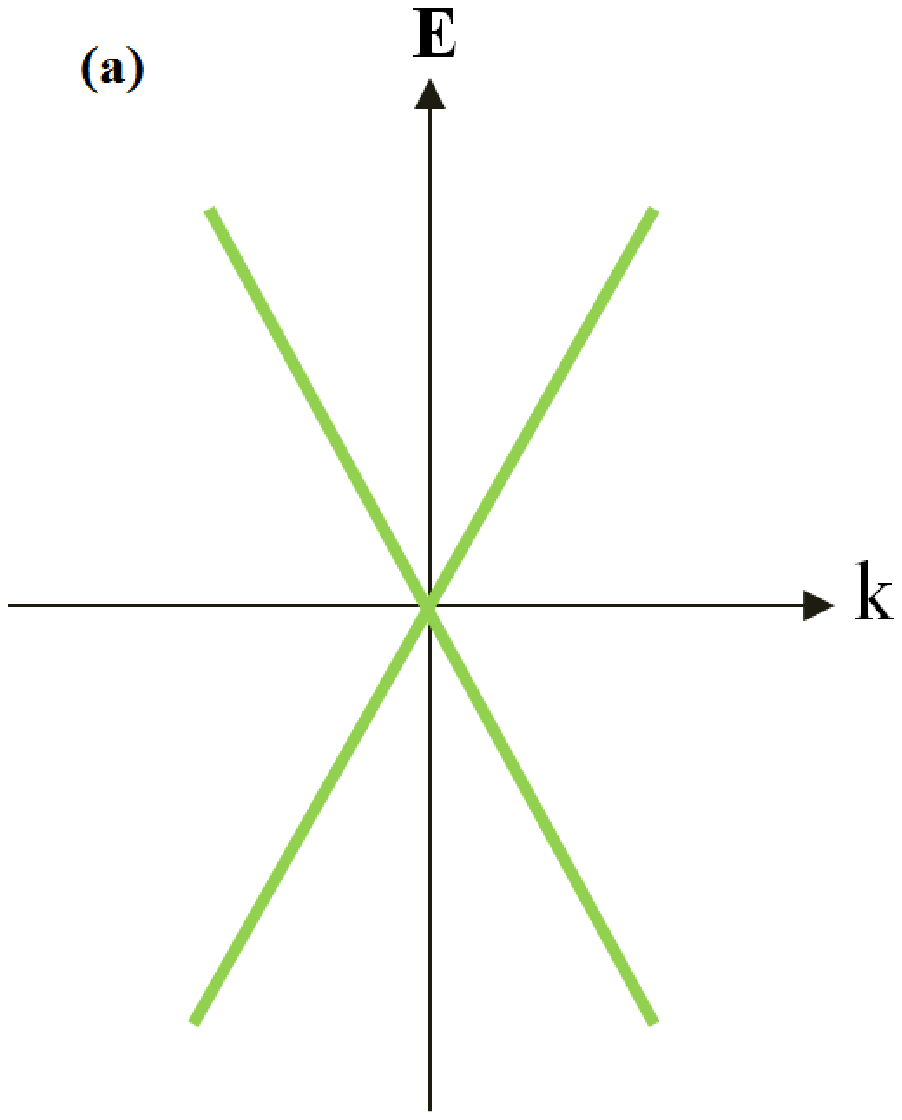} \
 \includegraphics[width=4.3cm , height=4.2cm]{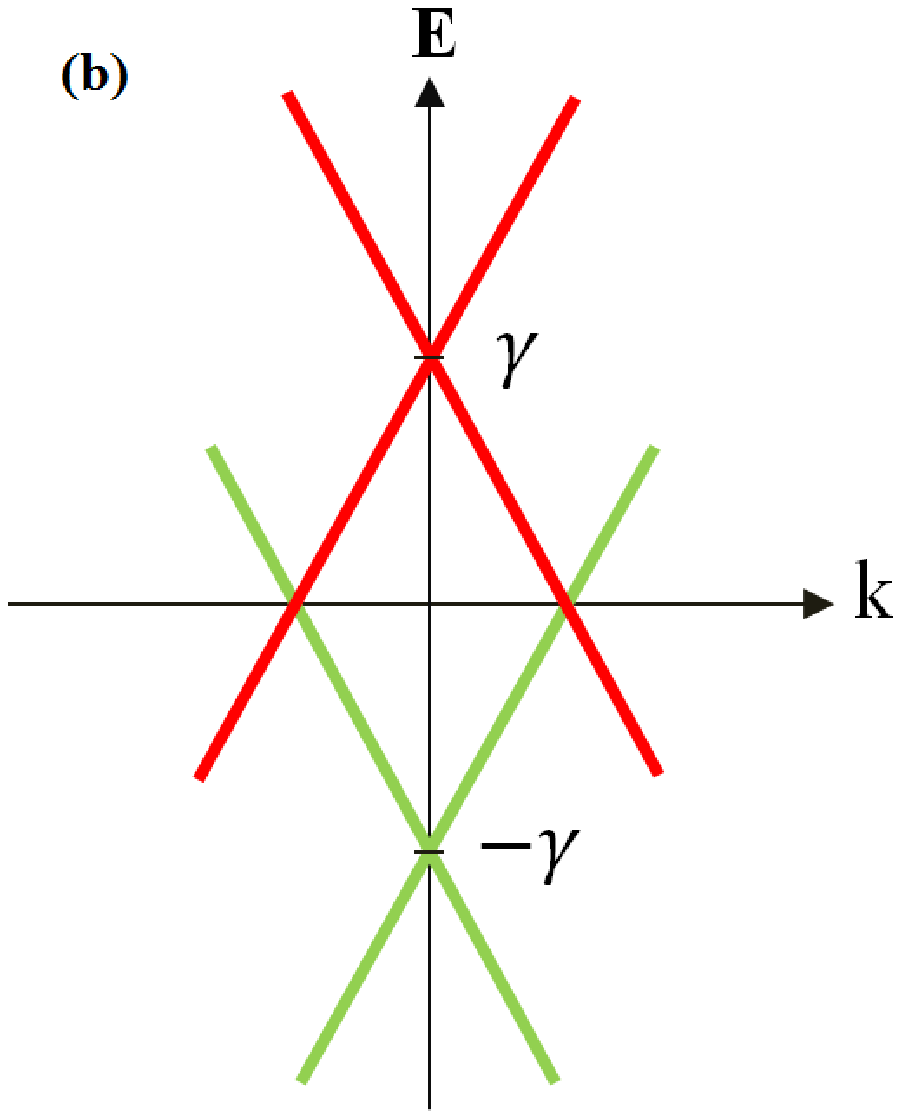}  \
 \includegraphics[width=4.1cm , height=4.2cm]{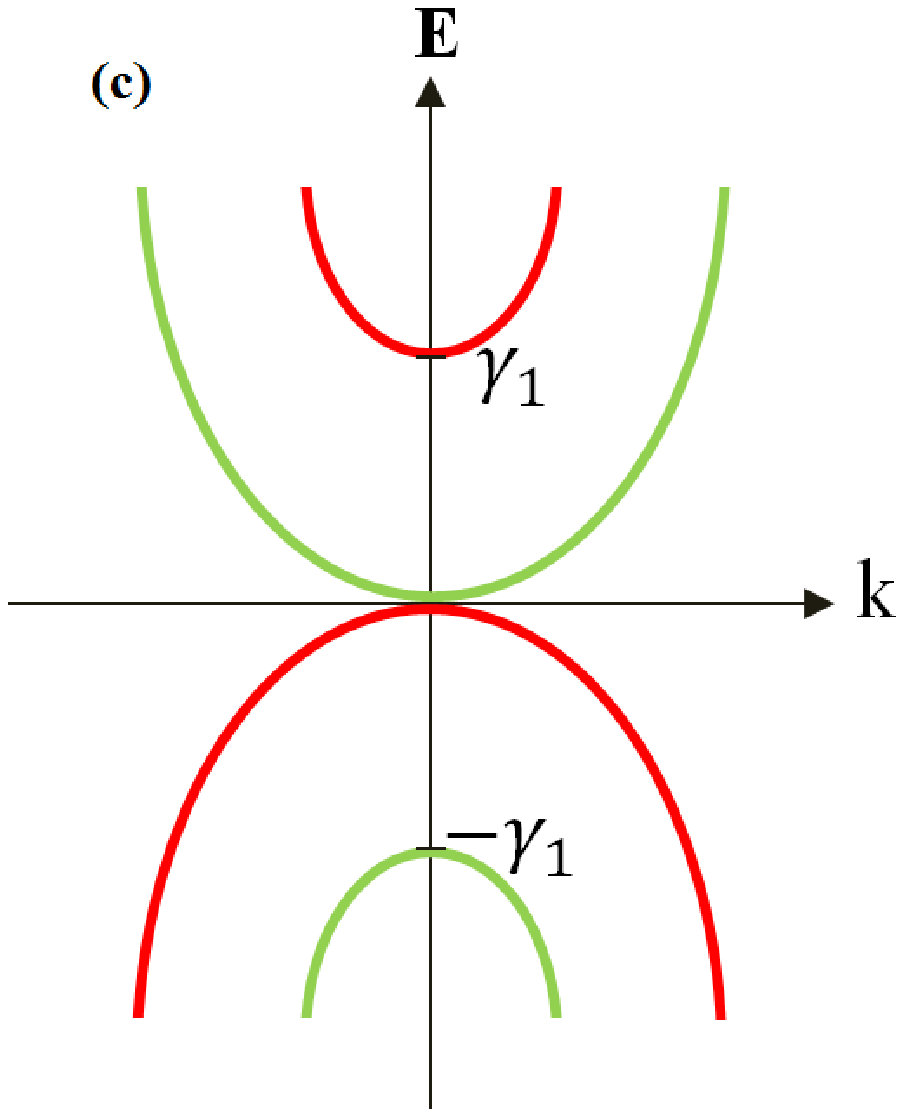}
  \caption{\sf{Energy bands of monolayer graphene (a), AA- stacked bilayer graphene (b) and AB-stacked bilayer graphene (c).
  In Figures (b) and (c), the green and
red lines correspond to the lower $(c=-1)$ and upper $(c=1)$ layers,
respectively.}}\label{BE}
\end{figure}

Let us in the beginning investigate the situation where the magnetic filed is switched off and underline
the behavior of our system.
Indeed, \eqref{DE3} for $V=0$ 
gives
\begin{subequations}\lb{DE-D0}
\beqar
&&\left(\frac{\partial}{\partial r}+\frac{m}{r}\right)\Phi_{A}(r)=-E\Phi_B(r)+ \gamma\Phi_{B'}(r) \\
&&\left(\frac{\partial}{\partial r}-\frac{m-1}{r}\right)\Phi_{B}(r)=E\Phi_A(r)- \gamma\Phi_{A'}(r)\\
&&\left(\frac{\partial}{\partial
r}-\frac{m-1}{r}\right)\Phi_{B'}(r)=E\Phi_{A'}(r)- \gamma\Phi_{A}(r)\\
&&\left(\frac{\partial}{\partial
r}+\frac{m}{r}\right)\Phi_{A'}(r)=-E\Phi_{B'}(r)+
\gamma\Phi_{B}(r). \eeqar
\end{subequations}
After some straightforward algebra, we end up with the Bessel differential
equation {for}
the component $\Phi_A(r)$. This is 
\beq
\lb{DEB} \left[r^2 \frac{\partial^2}{\partial r^2}+r
\frac{\partial}{\partial r}+ k^2 r^2 - m^2 \right]\Phi_A(r)=0.
\eeq
Note that the energy $E$ of ours system is related to the wave vector $k$ according to the
relation $k=s E + \nu \gamma$, where $s=\pm 1$ and $\nu =\pm 1$
are the band indices. Then, we define  the quantities $h \equiv s
$ and $c \equiv s \nu$ to write $k$ as
\beq\lb{khc}
k_{h,c}=h(E + c \gamma)
\eeq
with $h$ and $c$ will
be referred to the chirality index and the cone index,
respectively. Additionally, a quasiparticle state is situated in
the upper (lower) layer if $c = +1$ ($c = -1$). Hence, a
quasiparticle state is electron like (hole like) if $h = +1$ ($h =
-1$). This can be explained physically if we consider the group
velocity~\cite{AABZC10}. In fact, for $h = +1$ ($h=-1$), the wave
vector is parallel (anti parallel) to the group velocity. Note that
the energy spectrum for monolayer graphene
is linear~\cite{W47} and for 
the AA-stacked bilayer graphene
the energy bands are double copies of single layer
graphene bands shifted up and down by $\gamma$, respectively.
However, unlike monolayer and AA-stacked bilayer graphene, the
energy spectrum of the AB-stacked bilayer graphene are
parabolic~\cite{BP06} (Figure~\ref{BE}).

\noindent
The differential equation~\eqref{DEB} offers straightforward
solutions for $\Phi_A(r)$ in terms of the Bessel function, which is
\beq
\Phi_A(r)=C_1J_m(k_{h,+}r)+C_2J_m(k_{h,-}r)
\eeq
where the sign $+$
($c=+1$) and sign $-$ 
($c=-1$) correspond to the upper and lower layer, respectively.
While, the remains solutions can be obtained from \eqref{DE-D0} as
\beqar
&&\Phi_{B}(r) = -h\left[C_1J_{m-1}(k_{h,+}r)+C_2  J_{m-1}(k_{h,-}r)\right]  \nonumber \\
&&\Phi_{B'}(r) = h\left[C_1  J_{m-1}(k_{h,+}r)-C_2  J_{m-1}(k_{h,-}r)\right] \nonumber  \\
&&\Phi_{A'}(r) = -C_1J_m(k_{h,+}r) + C_2J_m(k_{h,-}r).
\eeqar

After obtaining the solution inside the quantum dot, it is natural to ask
about the eigenvalues associated to the bound states of the
system. To answer this inquiry we use the boundary conditions
given in 
\eqref{IMBC} to end up with
\beqar \label{EQ1}&&C_1\left[J_{m}(k_{h,+})-h \tau
J_{m-1}(k_{h,+})\right]=
C_2\left[ J_{m}(k_{h,-})-h \tau J_{m-1}(k_{h,-})\right]\\
\label{EQ2}
&&C_1\left[J_{m}(k_{h,+})+h \tau  J_{m-1}(k_{h,+})\right]=-
C_2\left[J_{m}(k_{h,-})+h \tau J_{m-1}(k_{h,-})\right].
\eeqar
These can be solved to obtain
the following characteristic equation for the allowed
eigenenergies $E$ of the quantum dot
\begin{equation}\label{eq65}
\frac{J_{m}(k_{h,+})-h \tau J_{m-1}(k_{h,+})}{J_{m}(k_{h,+})+h
\tau  J_{m-1}(k_{h,+})}+\frac{J_{m}(k_{h,-})-h \tau
J_{m-1}(k_{h,-})}{J_{m}(k_{h,-})+h \tau J_{m-1}(k_{h,-})}=0
\end{equation}
which will be numerically analyzed 
as well as compared to those obtained for monolayer and AB-staked
bilayer graphene systems.

\subsection{Nonzero magnetic field}

Now we consider our system in the presence of a perpendicular
magnetic field
and look for the solutions of the energy spectrum.
To proceed, we
choose the symmetric gauge $\vec{A}=\frac{B_0}{2}(0,r,0)$ to write the corresponding
momentum operators $\pi$ and $\pi^\dag$ as 
\beqar &&\pi=v_Fe^{i\theta}\left[-i\hbar (\frac{\partial}{\partial
r}+\frac{i\partial}{r\partial \theta})+i\frac{eB r}{2}\right]\\
&&\pi^\dag=v_Fe^{-i\theta}\left[-i\hbar (\frac{\partial}{\partial
r}-\frac{i\partial}{r\partial \theta})-i\frac{eB r}{2}\right].
\eeqar
Implementing these operators in the Hamiltonian~\eqref{eq42} and acting on the four-component wave
function $\Psi=e^{i m \theta} \left[\Phi_A,i e^{-i \theta
}\Phi_B,i e^{-i \theta }\Phi_{B'},\Phi_{A'} \right]^T$ 
to get
the four coupled first-order differential equations
\begin{subequations}\lb{eq68}
\beqar
&&\left(\frac{\partial}{\partial r}-\frac{m-1}{r}-\beta r\right)\Phi_{B}(r)=E\Phi_A(r)- \gamma\Phi_{A'}(r) \\
&&\left(\frac{\partial}{\partial r}+\frac{m}{r}+\beta r\right)\Phi_{A}(r)=-E\Phi_{B}(r)+ \gamma\Phi_{B'}(r) \\
&&\left(\frac{\partial}{\partial r}+\frac{m}{r}+\beta r\right)\Phi_{A'}(r)=-E\Phi_{B'}(r)+ \gamma\Phi_{B}(r) \\
&&\left(\frac{\partial}{\partial r} -\frac{m-1}{r}-\beta
r\right)\Phi_{B'}(r)=E\Phi_{A'}(r)- \gamma\Phi_{A}(r).
\eeqar
\end{subequations}
Decoupling the above equations with respect to $\Phi_{A}$ to
obtain 
\begin{equation}
\label{eq69} \left[\frac{\partial^2}{\partial
r^2}+\frac{1}{r}\frac{\partial}{\partial r}-\left( 2\beta (m-1)+
\frac{m^2}{r^2} + \beta^2 r^2- k_{h,c}^2\right)
\right]\Phi_{A}(r)= 0
\end{equation}
where $k_{h,c}$ is given in \eqref{khc} and $\beta=\frac{e B R}{2
\hbar}$. In order to solve the differential equation \eqref{eq69},
we make the following ansatz
\beq \label{eq70}
\Phi_{A}(r)=r^{|m|}e^{-\frac{r^2 \beta}{2}}\chi(r^{2})
\eeq
and
define a new variable
$\xi=\beta r^{2}$.
Substituting \eqref{eq70} into~\eqref{eq69} to get
\beq \label{eq71}
\left[\xi \frac{\partial^2}{\partial {\xi}^2
}+(b-\xi)\frac{\partial}{\partial \xi}-n_{h,c}\right]\chi(\xi)=0
\eeq
where we have set the quantities
\beq
b=1+|m|, \qquad n_{h,c}=-\frac{k_{h,c}^2 }{4
\beta}+\frac{m+|m|}{2}.
\eeq
\eqref{eq71} can be solved to get 
the following
combination
\begin{equation}
\label{eq72} \chi(\xi)=C_1
\tilde{M}(n_{h,+},b,\xi)+C_2\tilde{M}(n_{h,-},b,\xi)
\end{equation}
where the first term corresponds to the 
upper layer solution
and the second one corresponds to the 
lower layer solution. $\tilde{M}(n,b,\xi)$ is the regularized
confluent hypergeometric function, $C1$ and $C_2$ are
normalization constants.
From 
\eqref{eq70} and~\eqref{eq72},
we obtain the first spinor component \beq \label{eq73}
\Phi_A(r)=r^{|m|}e^{-\frac{\beta r^2}{2}}\left(C_1
\tilde{M}(n_{h,+},b,\beta r^2)+C_2 \tilde{M}(n_{h,-},b,\beta
r^2)\right) \eeq
and the others come as
\begin{eqnarray}
\Phi_B(r)&=&r^{|m|-1}e^{-\frac{\beta r^2}{2}}\left[
\frac{C_1}{2(\gamma + E)} \left(r^2\eta_+
\tilde{M}(n_{h,+},b+1,\beta r^2)  - \zeta
\tilde{M}(n_{h,+},b,\beta r^2)\right) \right. \nonumber \\
&&\left.
 -\frac{C_2}{2(\gamma - E)}\left(r^2\eta_-\tilde{M}(n_{h,-},b+1,\beta r^2)- \zeta
 \tilde{M}(n_{h,-},b,\beta r^2) \right)\right]\\
\Phi_{B'}(r)&=&-r^{|m|-1}e^{-\frac{\beta r^2}{2}}\left[
\frac{C_1}{2(\gamma + E)}\left(r^2\eta_+
\tilde{M}(n_{h,+},b+1,\beta r^2)  - \zeta
\tilde{M}(n_{h,+},b,\beta r^2)\right) \right. \nonumber \\
&&\left. +\frac{C_2}{2(\gamma -
E)}\left(r^2\eta_-\tilde{M}(n_{h,-},b+1,\beta r^2)- \zeta
 \tilde{M}(n_{h,-},b,\beta r^2) \right)\right]\\
\label{eq76} \Phi_{A'}(r)&=&-r^{|m|}e^{-\frac{\beta
r^2}{2}}\left(C_1 \tilde{M}(n_{h,+},b,\beta r^2)-C_2
\tilde{M}(n_{h,-},b,\beta r^2)\right)
\end{eqnarray}
where the quantities are 
\beq
\eta_{\pm}=2\beta(|m|-m+2)+(\gamma \pm E)^2, \qquad
\zeta=2(|m|+m+2 \beta r^2).
\eeq

To get the allowed energy for the quantum dots we use again
the infinite-mass boundary condition given in
\eqref{IMBC}. Doing this to get 
\beqar \label{eq77}
&&C_{1}\left[\left(1-\frac{\tau \zeta}{2(\gamma +E)}\right)
\tilde{M}(n_{h,+},b,\beta)+\frac{\tau \eta_+
}{2(\gamma +E)} \tilde{M}(n_{h,+},b+1,\beta) \right] \nonumber\\
&&= C_{2}\left[\left( 1+\frac{\tau \zeta}{2(\gamma -E)}\right)
\tilde{M}(n_{h,-},b,\beta)-\frac{\tau \eta_-}{2(\gamma -E)}
\tilde{M}(n_{h,-},b+1,\beta)\right]
\eeqar
\begin{eqnarray}
\label{eq78} 
&&C_{1}\left[\left(1+\frac{\tau \zeta}{2(\gamma
+E)}\right)\tilde{M}(n_{h,+},b,\beta)-\frac{\tau \eta_+}{2(\gamma
+E)} \tilde{M}(n_{h,+},b+1,\beta) \right] \nonumber\\
&&= -C_{2}\left[ \left( 1-\frac{\tau \zeta}{2(\gamma
-E)}\right)\tilde{M}(n_{h,-},b,\beta)+\frac{\tau \eta_-}{2(\gamma
-E)} \tilde{M}(n_{h,-},b+1,\beta)\right].
\end{eqnarray}
By the eliminating the constants $C_1$ and $C_2$, we finally  
find the characteristic equation
for the allowed eigenenergies $E$ in the case of nonzero magnetic field. This is
\begin{eqnarray}
\label{eq79} &&\frac{(2(\gamma+E)-\tau
\zeta)\tilde{M}(n_{h,+},b,\beta)+\tau \eta_+
\tilde{M}(n_{h,+},b+1,\beta)}{(2(\gamma+E)+\tau
\zeta)\tilde{M}(n_{h,+},b,\beta)-\tau \eta_+
\tilde{M}(n_{h,+},b+1,\beta)} \nonumber \\&& +
\frac{(2(\gamma-E)+\tau \zeta )\tilde{M}(n_{h,-},b,\beta)-\tau
\eta_- \tilde{M}(n_{h,-},b+1,\beta)}{(2(\gamma-E)-\tau
\zeta)\tilde{M}(n_{h,-},b,\beta)+\tau \eta_-
\tilde{M}(n_{h,-},b+1,\beta)}=0.
\end{eqnarray}

After obtaining the infinite-mass boundary condition and the
eigenenergies of the QD in the presence {and
absence} of a perpendicular magnetic field, we numerically study
the energy levels by using the
above equations and underline the behavior of our system. 

\section{
Results and discussion} \lb{Rd}

Let us first discuss the energy spectrum in the absence of magnetic field
by using equation 
\eqref{eq65}. In discussing
the properties of the energy spectrum, we use the notation
$E^{q}(\tau,m)$, where $q \equiv e\ (h)$ denotes the electron
(hole), $\tau$ is the valley index and it is equal to $+1$ ($-1$)
for the $K$ ($K'$) valley and $m$ correspond to the total angular
quantum number. The fact that the Bessel functions obey the
following properties
\beq J_{n}(x)=(-1)^n J_{-n}(x),\qquad
J_{n}(x)=(-1)^n J_{n}(-x)
\eeq
we can derive interesting
results. First, the intervalley spectrum symmetry is present,
where the $m$ states in the $K$ valley have equal energies to the
$-m+1$ states in the $K'$ valley
\beq \lb{39}
E^{e,h}(\tau, m)=E^{e,h}(-\tau, -m+1)
\eeq
Furthermore, there is  a
symmetry between the electron $m$ states and the hole $-m+1$
states
\beq \lb{40} E^{e}(\tau, m)=-E^{h}(\tau,
-m+1)
\eeq
Also, we have another interesting symmetry indicate the
intervalley electron-hole symmetry between the states of the same
quantum number $m$ 
\beq \lb{41}E^{e}(\tau,
m)=-E^{h}(-\tau, m)
\eeq
The intervalley electron-hole symmetry are
also present in the case for AB-stacked bilayer
graphene~\cite{DMAGF14} with infinite-mass
barrier. 
We note that our results are compatible with those obtained for
monolayer graphene for infinite-mass boundary
conditions~\cite{GZCTFP11}.

\begin{figure}[!ht]
  \centering
  \includegraphics[width=5.4cm, height=5cm]{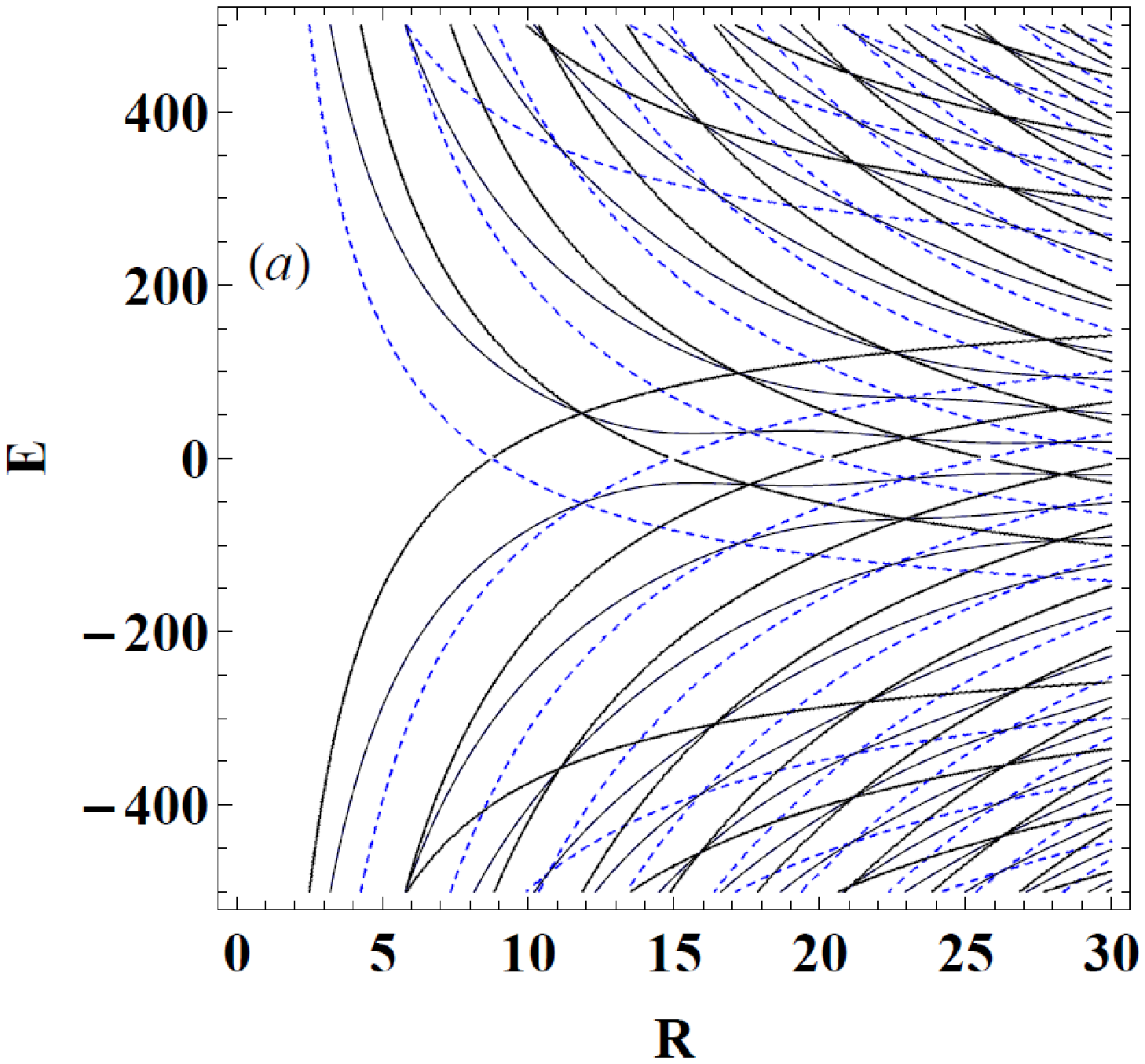}  \includegraphics[width=5.4cm, height=5cm]{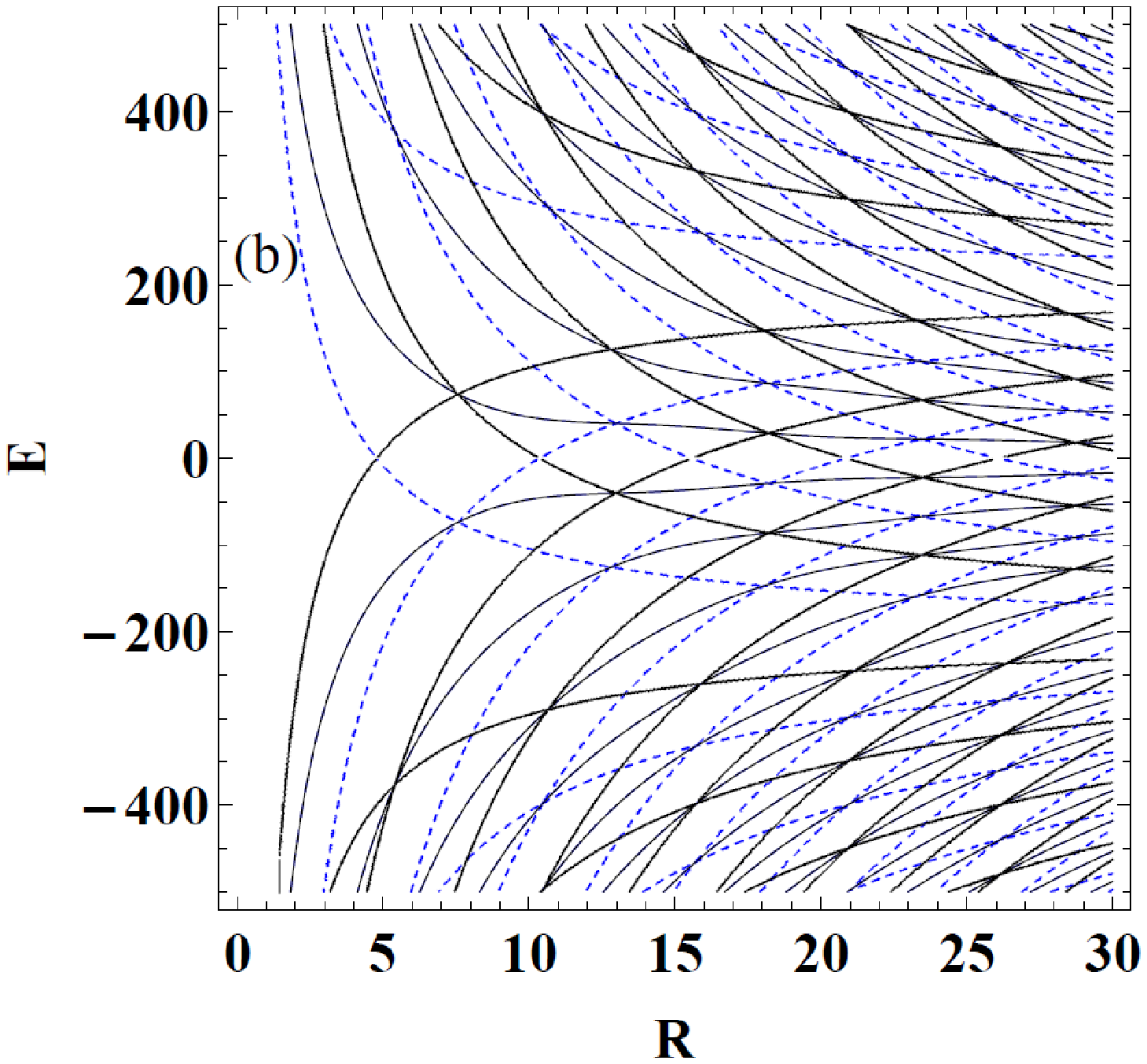} \
  \includegraphics[width=5.4cm , height=5cm]{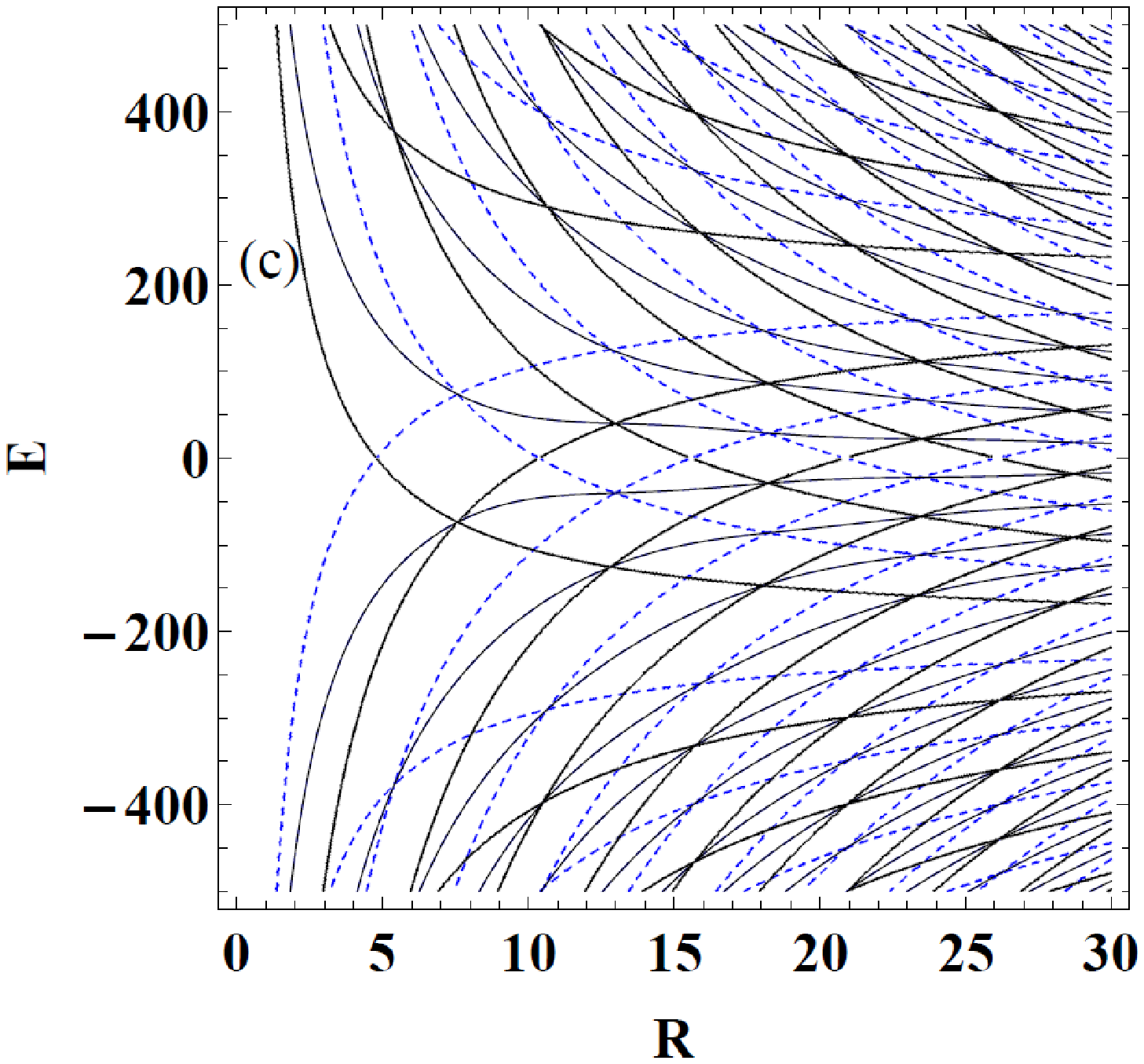}
  \caption{\sf{Energy spectrum of a circular AA-stacked bilayer graphene quantum dots as a function of the dot
  radius. (a) $m=-1$, (b) $m=0$ and (c) $m=1$. The energy states for the $K$ valley ($\tau=1$) are drown with dashed blue lines, those for the
   $K'$ valley ($\tau=-1$) with solid black lines.}}\label{ESB0}
\end{figure}

In Figure~\ref{ESB0}, we plot the energy levels for a circular
AA-stacked bilayer graphene QD as function of the dot radius $R$
for $m=\pm 1$ and $m=0$. The $K$ valley spectrum is depicted by
the solid black lines, while the dashed blue lines denote the
energy levels in the $K'$ valley. One can see that the energy
spectrum shows two set of levels. They are just the double copies
of the energy spectrum corresponding to single layer graphene, one
shifted up by $+\gamma$ and 
other one shifted down by $-\gamma$, where $\gamma=200\ meV$ is
the interlayer coupling. We notice that the upper one corresponds
to the upper layer and the lower one corresponds to the lower
layer.
For large $R$, the set of levels corresponding to the upper layer
converges to the interlayer
hopping energy $\gamma=200\ meV$. However, the set levels corresponding to
the lower layer, converge to $-200\ meV$.

\begin{figure}[!ht]
  \centering
  \includegraphics[width=5.4cm, height=5cm]{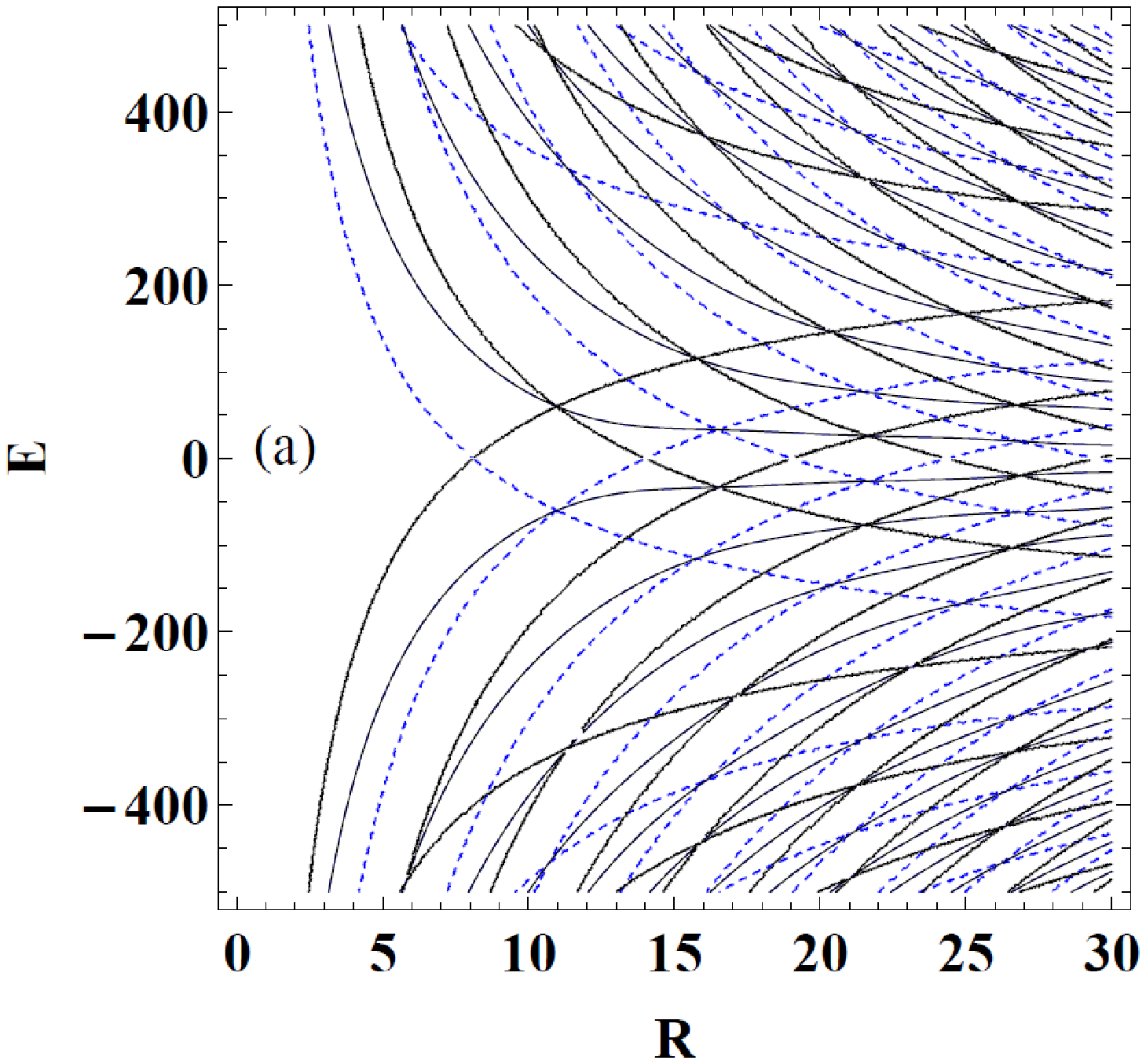}  \includegraphics[width=5.4cm, height=5cm]{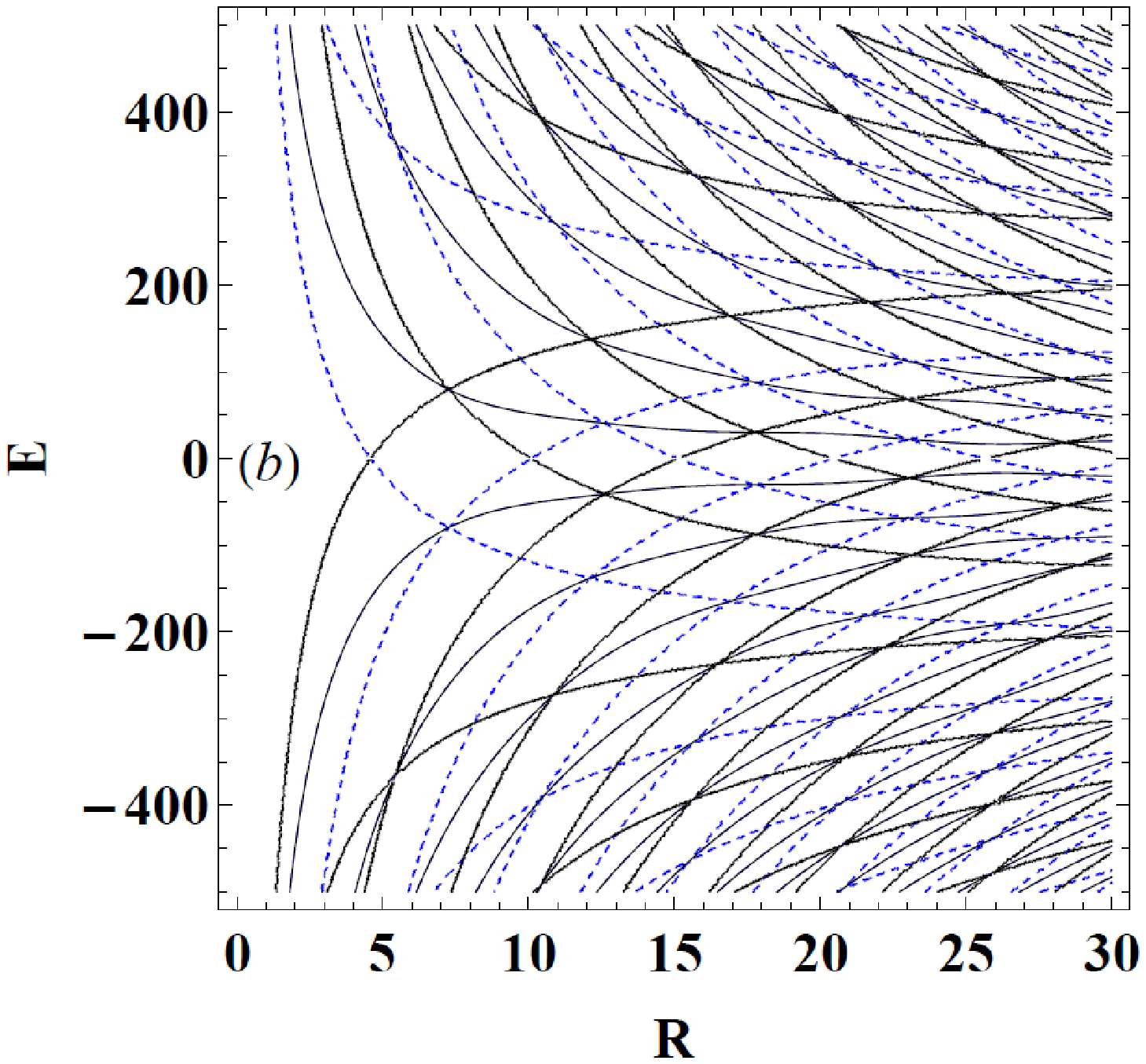} \
  \includegraphics[width=5.4cm , height=5cm]{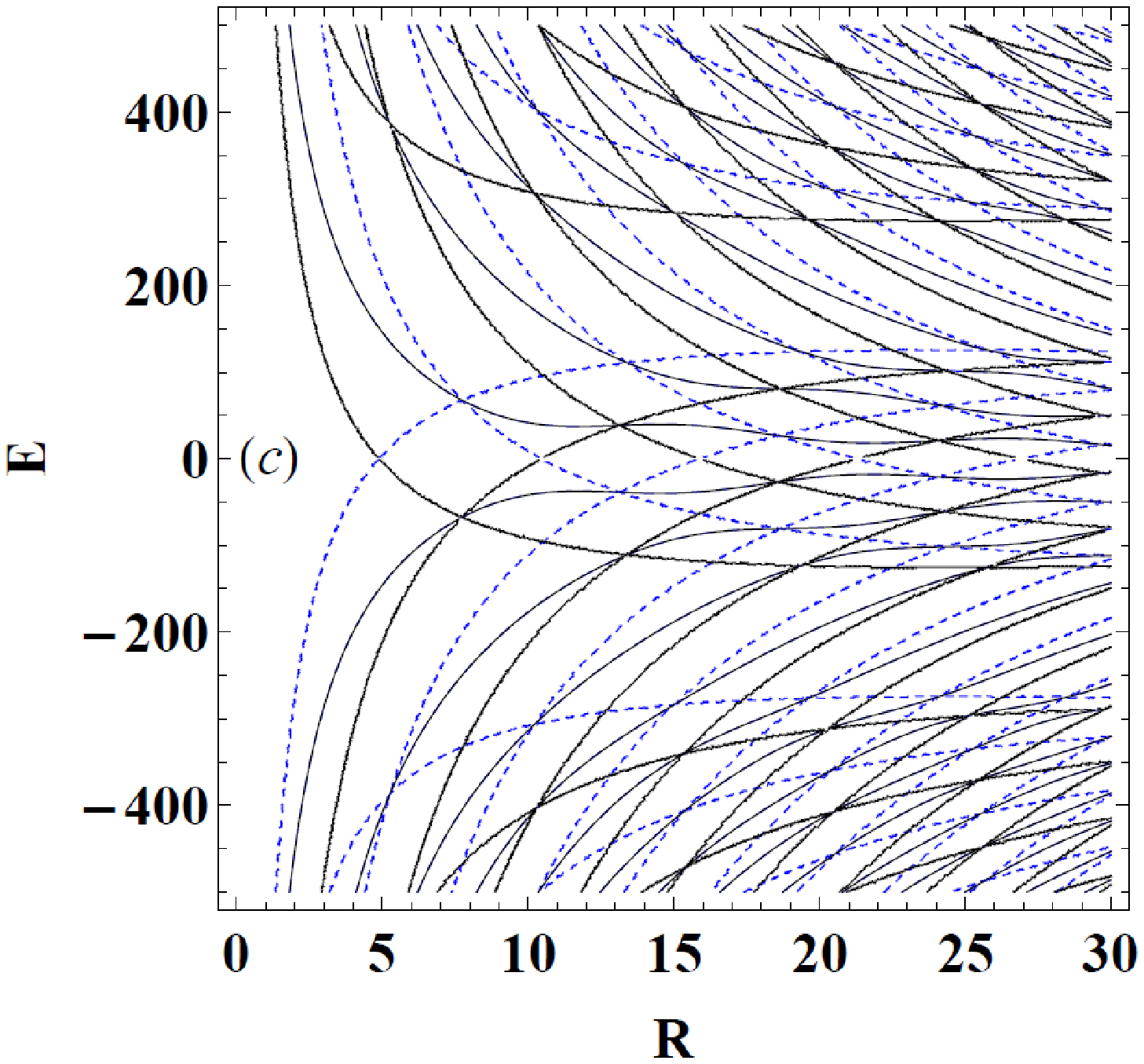}
  \caption{\sf{Energy spectrum of a circular AA-stacked bilayer graphene quantum dots as a function of the dot
  radius. (a) $m=-1$, (b) $m=0$ and (c) $m=1$ with $B=5\ T$. The energy
states for the $K$ valley ($\tau=1$) are drown with dashed blue
lines. Those for the
   $K'$ valley ($\tau=-1$) with solid black lines.}}\label{ESB}
\end{figure}

Now, we switch to the case with a nonzero magnetic field. In
Figure~\ref{ESB}, we plot the energy levels as function of the dot
radius $R$ for $m=0,\ \pm 1$ and $B=5T$ for both valleys. From
\eqref{eq79} we notice that the presence of the magnetic
field breaks all symmetry properties except one
\beq
E^{e}(\tau, m)=-E^{h}(-\tau, m)
\eeq
We
clearly observe form Figures~\ref{ESB0} and \ref{ESB}, the
crossing between the levels of two different valleys for $E=0$.
Also, as long as the size of the dot is increasing the energy spectrum becomes
weakly dependent on the dot radius. In addition, we notice that this
symmetry is obtained for monolayer~\cite{GZCTFP11} and AB-stacked
bilayer graphene~\cite{DMAGF14} in the presence of nonzero
magnetic field with infinite-mass barrier.

\begin{figure}[h!]
  \centering
  \includegraphics[width=7.6cm, height=7.7cm]{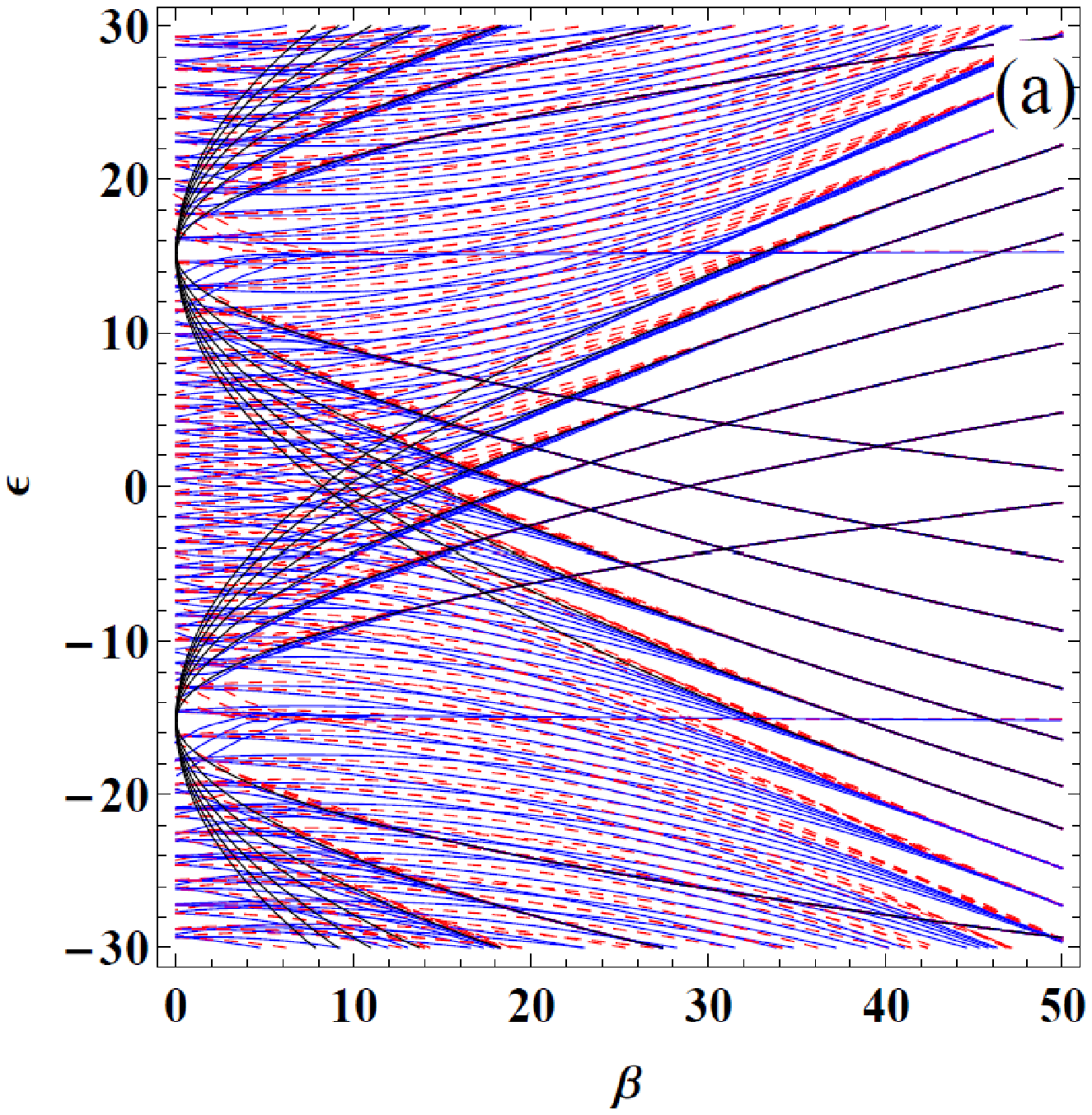}
 \ \ \includegraphics[width=7.6cm, height=7.7cm]{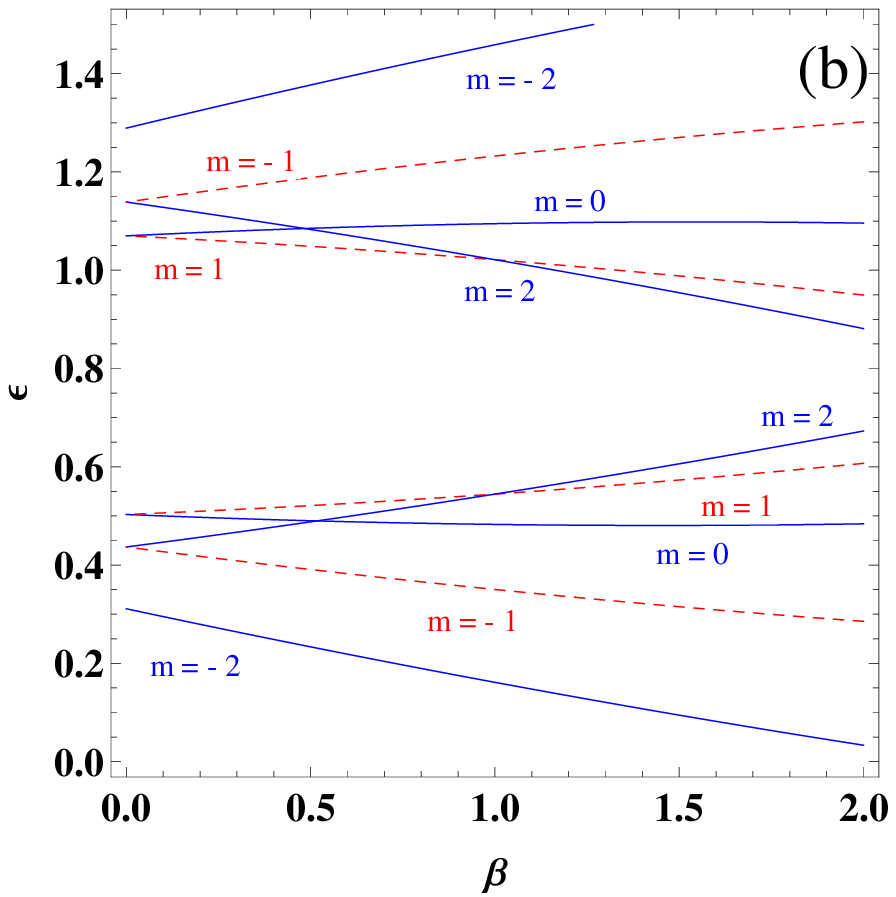}
  \caption{\sf{Energy spectrum of a circular AA-stacked bilayer graphene quantum dots as a function of
 a perpendicular magnetic field for $R=50\ nm$ and $-2 \leq m \leq 2$. The energy states for the $K$ valley ($\tau=1$) are drown with
solid blue lines, those for the $K'$ valley ($\tau=-1$) with
dashed red lines. The black solid lines correspond the Landau
levels of AA-stacked bilayer graphene.}}\label{E-B}
\end{figure}

Now we study the energy
spectrum versus the  magnetic field. In Figure~\ref{E-B}, we show the energy levels of a
circular quantum dot as a function of a perpendicular magnetic
field for $R=50\ nm$ and $-2 \leq m \leq 2$. The energy levels
corresponding to the $K$ and $K'$ valleys are shown, respectively,
by the solid blue lines and dashed red lines.
We can clearly see from
Figures~\ref{ESB} that for zero magnetic field the energy states
are not equidistant in contrast to the case of semiconductor
QDs~\cite{L91}. From Figure~\ref{E-B}(a), we observe that
for higher magnetic field the energy levels approach the Landau
levels (LLs) of AA-stacked bilayer graphene.
Moreover, the carrier confinement can be subdivided in two
regimes. The first one is characterized by the weak magnetic field, where the
confinement is due to the infinite-mass barrier. Whereas, the
second regime is manifested at large magnetic field where the
influence of the infinite-mass barrier is suppressed and the
confinement becomes dominated by the magnetic field. We notice that the
transition between these two confinement regimes takes place as the
magnetic field increases. The transition points between the two
regimes can be defined when the energies of the states in the
dot differ negligibly from the LLs energy. It can be seen that
the transition points shift toward larger magnetic field with
lower $m$.

\noindent Figure~\ref{E-B}(b) shows an enlargement of the low energy levels,
corresponding to both valleys $K$ and $K'$, at small magnetic
field. From this 
we can see that, for zero magnetic field, the two degenerate
states $E(1,m=2)$ and $E(-1,m=-1)$ belong to different valleys.
Also, the others degenerate levels $E(1,m=0)$ and $E(-1,m=1)$
belong to different valleys. This result confirms the intervalley
spectrum symmetry $ E^{e,h}(\tau,m) =
E^{e,h}(-\tau,-m+1)$. It may be noted that the application of a
perpendicular magnetic field leads to the discretization of the
energy spectrum. This can explain the fact that for nonzero
magnetic field the degeneracy is broken.

In summary, we notice that the asymptotic behavior of the energy
levels for both valleys $K$ and $K'$ can be derived 
by using infinite-mass boundary conditions for large magnetic field. Doing this
to end up with
\beq \lb{LL}
E^{e,h}_{m,n}(\beta)=c \gamma \pm \sqrt{4
\beta \left(n+\frac{m+|m|}{2} \right)}
\eeq
where $c=\pm 1$ refer
to the cone index, the sign $\pm$ accounts for the particle type
(electron/hole) and the Landau levels index is denoted by $n$.
Hence, we retrieve the well-known LLs for monolayer graphene from
\eqref{LL} for the limit $\gamma = 0$.
The first LL ($n=1$) is composed of $m\leq 0$ in both valleys $K$
and $K'$. Similarly, the higher energy LLs ($n>1$) satisfy the
condition $m\leq n$. This behavior is similar to that in
monolayer~\cite{GZCTFP11} and AB-stacked bilayer
graphene~\cite{DMAGF14} with infinite-mass barrier and also for
semiconductors~\cite{L91}.

\section{Conclusion}

 In this paper, we considered a system of bilayer of graphene in the
 Bernal AA stacking and assumed that the carriers are confined in a circular areas of
radius $R$, which was modeled by an infinite-mass barrier. From the eigenvalue equation, we
derived
the infinite-mass boundary condition in AA-stacked bilayer
graphene quantum dot. This was only region I dependent, which
means that the region outside the dot was forbidden for particles.
The {infinite-mass} boundary condition allowed us
to
investigate the energy levels 
in the absence and  presence  of the magnetic field.

More precisely, after obtaining the corresponding wave functions
inside and outside the quantum dot, we applied the obtained
{infinite-mass} boundary conditions
to find the energy levels of a circular graphene quantum dots for
zero and nonzero magnetic field. The energy spectrum
exhibited an intervalley spectrum symmetry (\eqref{39},
\eqref{41})
and also an electron-hole symmetry  \eqref{40}.
Furthermore, we found that the presence of a nonzero magnetic field
broke all symmetry properties except one given in \eqref{41}. The
obtained energy spectrum symmetry are similar to those obtained in
monolayer graphene for infinite-mass boundary conditions. However,
the intervalley electron-hole symmetry is also present in the
case of AB-stacked bilayer graphene.

Our results showed that the energy spectrum presents two sets of
states as function of the quantum dot radius $R$. The upper set corresponds to
the upper layer and the lower one corresponds the the lower layer.
For large $R$, the upper set converges to the interlayer hopping
energy $+\gamma$, whereas the lower set converges to $-\gamma$.
In addition, we analyzed the magnetic field dependence of the
energy spectrum. Indeed, for zero magnetic field, we found that the energy
states are not equidistant in contrast to the results reported  for
semiconductor quantum dots. However, in the presence of a magnetic
field we obtained two regimes of carriers confinement: for weak
magnetic field the confinement is due to the infinite-mass
barrier and
for large
magnetic field the influence of the infinite-mass barrier is
suppressed, hence the confinement becomes dominated by magnetic
field. Therefore, the transition between these two confinement
regimes takes place as long as the magnetic field increases.
By increasing the magnetic field,
the energy levels of a circular quantum dot approach the Landau
levels for AA-stacked bilayer graphene. Our results showed
that the degenerate states, for zero magnetic field, belong to
different valleys. For nonzero magnetic field, this degenerate was
broken, owing to the discretization of the energy spectrum when a
magnetic field is applied.

\section*{Acknowledgment}

The generous support provided by the Saudi Center for Theoretical
Physics (SCTP) is highly appreciated by all authors.

\end{document}